%% file: DE_QDs_Nanoscale_for_arXiv.tex
\documentclass[
floatfix, %
twocolumn, %
reprint, %
superscriptaddress, %
prb, %
aps, %
nofootinbib,%
nobibnotes,%
longbibliography, %
lengthcheck, %
final, %
]{revtex4} 
\renewcommand{\thispagestyle}[1]{}
\usepackage[T1]{fontenc}
\usepackage[utf8]{inputenc}
\usepackage{graphicx}
\usepackage{latexsym}
\usepackage[low-sup]{subdepth}
\usepackage{soul}
\usepackage{color}
\usepackage{mathtools}
\usepackage{ulem}
\usepackage{notes2bib}

\usepackage[]{placeins}

\usepackage[dvipsnames]{xcolor}

\usepackage{upgreek}

\usepackage{bm}
\usepackage[scale=0.97]{XCharter} 
\usepackage[libertine,bigdelims,vvarbb,scaled=1.05]{newtxmath} 
 
\usepackage[babel=true]{microtype}

\usepackage{floatrow}
\usepackage{siunitx}
\mathchardef\mhyphen="2D
\usepackage{hyperref}
\hypersetup{
	colorlinks,
	linkcolor={blue!90!black},
	citecolor={blue!90!black},
	urlcolor=	{blue!90!black}
}
\let\oldeqref\eqref
\renewcommand*{\eqref}[1]{%
	\hyperref[#1]{\oldeqref{#1}}%
}

\usepackage{ragged2e}
\usepackage{multirow}
\usepackage[flushleft]{threeparttable} %
\usepackage{array}

\newcommand*{\kp}{\bm{k}{\cdot}\bm{p}}
\newcommand{\eqnref}[1]{Eq.~\eqref{eq:#1}}
\newcommand{\tabref}[1]{Tab.~\ref{tab:#1}}

\newcommand{\figref}[1]{Fig.~\ref{fig:#1}}
\newcommand{\Supplfigref}[1]{Fig.~\ref{fig:#1}}

\newcommand{\subfigref}[2]{Fig.~\hyperref[fig:#1]{\ref*{fig:#1}#2}}
\newcommand{\Supplsubfigref}[2]{Fig.~\hyperref[fig:#1]{\ref*{fig:#1}#2}}
\newcommand{\subfigrefL}[2]{Figure~\hyperref[fig:#1]{\ref*{fig:#1}#2}}
\newcommand{\subfigsref}[3]{Figs.~\hyperref[fig:#1]{\ref*{fig:#1}#2}-\hyperref[fig:#1]{\ref*{fig:#1}#3}}
\newcommand{\Supplsubfigsref}[3]{Figs.~S\hyperref[fig:#1]{\ref*{fig:#1}#2}-S\hyperref[fig:#1]{\ref*{fig:#1}#3}}
\newcommand{\subfigsrefL}[3]{Figures~\hyperref[fig:#1]{\ref*{fig:#1}#2}-\hyperref[fig:#1]{\ref*{fig:#1}#3}}

\definecolor{cbred}{HTML}{e31a1c}
\definecolor{cbgreen}{HTML}{33a02c}
\definecolor{cbblue}{HTML}{176aa7}
\definecolor{cborange}{HTML}{ff7f00}
\definecolor{cbviolet}{HTML}{6a3d9a}

\definecolor{tomimoto}{HTML}{66c2a5}
\definecolor{gong}{HTML}{e78ac3}

\DeclarePairedDelimiterX{\comm}[2]{\lbrack}{\rbrack}{#1, #2}

\DeclarePairedDelimiterX{\braket}[2]{\langle}{\rangle}{#1\delimsize\vert #2}
\DeclarePairedDelimiterX{\ketbra}[2]{\rvert}{\lvert}{#1 \delimsize\rangle\!\delimsize\langle #2}
\DeclarePairedDelimiterX{\matrixel}[3]{\langle}{\rangle}{#1 \delimsize\vert #2 \delimsize\vert #3}

\begin{document}

\title{Droplet epitaxy symmetric InAs/InP quantum dots for quantum emission in the third telecom window: morphology, optical and electronic properties}
	
	\author{P.~Holewa}
	\email{pawel.holewa@pwr.edu.pl}
		\affiliation{
		DTU Fotonik, Technical University of Denmark, Kongens Lyngby DK-2800, Denmark
	}%
	\affiliation{%
		Laboratory for Optical Spectroscopy of Nanostructures, %
		Department of Experimental Physics, %
		Faculty of Fundamental Problems of Technology, %
		Wroc\l{}aw University of Science and Technology, %
		Wybrze\.ze Wyspia\'nskiego 27, %
		50-370 Wroc\l{}aw, %
		Poland%
	}

	\author{S.~Kadkhodazadeh}
	\affiliation{
		DTU Nanolab-National Centre for Nano Fabrication and Characterization, Technical University of Denmark, Kongens Lyngby DK-2800, Denmark
	}%
	\affiliation{NanoPhoton-Center for Nanophotonics, Technical University of Denmark, DK-2800 Kongens Lyngby, Denmark}
	
	\author{M. Gawe\l{}czyk}
	\affiliation{
Department of Theoretical Physics, Faculty of Fundamental Problems of Technology, Wrocław University of Science and Technology, 50-370 Wrocław, Poland	}%
	\affiliation{
Institute of Physics, Faculty of Physics, Astronomy and Informatics, Nicolaus Copernicus University, Toru\'n 87-100, Poland	}%
	
	\author{P.~Baluta}
	\affiliation{%
		Laboratory for Optical Spectroscopy of Nanostructures, %
		Department of Experimental Physics, %
		Faculty of Fundamental Problems of Technology, %
		Wroc\l{}aw University of Science and Technology, %
		Wybrze\.ze Wyspia\'nskiego 27, %
		50-370 Wroc\l{}aw, %
		Poland%
	}
	\author{A.~Musia\l{}}
	\affiliation{%
		Laboratory for Optical Spectroscopy of Nanostructures, %
		Department of Experimental Physics, %
		Faculty of Fundamental Problems of Technology, %
		Wroc\l{}aw University of Science and Technology, %
		Wybrze\.ze Wyspia\'nskiego 27, %
		50-370 Wroc\l{}aw, %
		Poland%
	}	
		\author{V. Dubrovskii}
	\affiliation{
Faculty of Physics, St. Petersburg State University, Universitetskaya Embankment 13B, 199034, St. Petersburg, Russia	}%

	\author{M.~Syperek} 
	\affiliation{%
		Laboratory for Optical Spectroscopy of Nanostructures, %
		Department of Experimental Physics, %
		Faculty of Fundamental Problems of Technology, %
		Wroc\l{}aw University of Science and Technology, %
		Wybrze\.ze Wyspia\'nskiego 27, %
		50-370 Wroc\l{}aw, %
		Poland%
	}
	\author{E.~Semenova} 
	\affiliation{
		DTU Fotonik, Technical University of Denmark, Kongens Lyngby DK-2800, Denmark
	}%
	\affiliation{NanoPhoton-Center for Nanophotonics, Technical University of Denmark, DK-2800 Kongens Lyngby, Denmark}

\begin{abstract}
The rapidly developing quantum communication technology requires deterministic quantum emitters that can generate single photons and entangled photon pairs in the third telecom window, in order to be compatible with existing optical fiber networks and on-chip silicon photonic processors. InAs/InP quantum dots (QDs) are among the leading candidates for this purpose, due to their high emission efficiency in the required spectral range. However, fabricating versatile InAs/InP QD-based quantum emitters is challenging, especially as these QDs typically have asymmetric profiles in the growth plane, resulting in a substantial bright-exciton fine structure splitting (FSS). This hinders the generation of entangled photon pairs and thus, compromises the versatility of InAs/InP QDs. We overcome this by implementing droplet epitaxy (DE) synthesis of low surface density ($\SI{2.8e8}{\centi\meter\tothe{-2}}$) InAs QDs on an $(001)$-oriented InP substrate. The resulting QDs are located in etched pits, have concave bases, and most importantly, have symmetric in-plane profiles. We provide an analytical model to explain the kinetics of pit formation and QD base shape modification. Our theoretical calculations of electronic states reveal the properties of neutral and charged excitons and biexcitons confined in such QDs, which agree with the optical investigations of individual QDs. The optical response of QD ensembles suggests that FSS may indeed be negligible, as reflected in the vanishing degree of linear polarization. However, single QD spectrum gathered from an etched mesa shows moderate FSS of $(50\pm5)~\si{\micro\electronvolt}$ that we link to destructive changes made in the QD environment during the post-growth processing. Finally, we show that the studied DE QDs provide a close-to-ideal single-photon emission purity of $(92.5\pm7.5)~\%$ in the third telecom window.
\end{abstract}
	
\maketitle
	
\section{Introduction}
Epitaxially-grown semiconductor quantum dots (QDs), among other applications~\cite{Aharonovich2016}, are considered as nearly perfect quantum emitters~\cite{Senellart2017,Schweickert2018} for applications in quantum communication~\cite{Basset2021} and quantum computation~\cite{Aharonovich2016,Wang2019NatPhot} technologies.
QDs emitting in the long-wavelength third telecom band centered at $\SI{1550}{\nano\meter}$ are of special interest for these applications~\cite{Cao2019}, as they offer ultra-low-loss single-photon-encoded data transmission between quantum nodes, both in distributed silica-fiber-based optical networks~\cite{Wehner2018,Cuomo2020} and on-chip optical circuits~\cite{Vlasov2004,Kim2017Emitter,Tran2018,Hepp2019,Elshaari2020}. Two main material systems have been identified for long-wavelength QD-based single-photon emission: InAs/GaAs~\cite{Semenova2008,Paul2017,Nawrath2019,Zeuner2019} and InAs/InP~\cite{Miyazawa2016,Skiba-Szymanska2017,Ha2020,Musial2019,Shikin2019,Holewa2020PRB,Holewa2020PRApplied}.
InAs/GaAs QDs show excellent single-photon emission properties~\cite{Portalupi2019}, including high single-photon emission purity~\cite{Paul2017,Zeuner2019}, and high degree of photon indistinguishability~\cite{Nawrath2019}. This, however, is at the cost of complicated strain engineering epitaxial processes to red-shift the emission~\cite{Cao2019,Arakawa2020}.
In comparison, InAs/InP QDs naturally emit at longer wavelengths, while also offering high photon emission purity~\cite{Miyazawa2016,Arakawa2020} and indistinguishability~\cite{Anderson2021}.
However, controlling their growth kinetics and the resulting morphology with respect to size, shape anisotropy, and surface density is complicated. Therefore, establishing an epitaxial method that allows tailoring the aforementioned mentioned properties independently will make InAs/InP QDs highly attractive as solid-state single-photon quantum emitters in the long-wavelength telecom range.

There are two main epitaxial techniques for synthesizing InAs/InP QDs: Stranski-Krastanov (SK)~\cite{Miyazawa2016,Musial2019,Holewa2021Mirror}, and droplet epitaxy (DE)~\cite{Skiba-Szymanska2017,Ha2020,Gurioli2019}.
SK InAs/InP QDs hold the record for single-photon purity in the long-wavelength spectral range~\cite{Miyazawa2016}, whereas DE InAs/InP QDs have the lead in entangled photon pair emission~\cite{Muller2018} due to their more symmetric shape compared to SK QDs~\cite{Skiba-Szymanska2017}. The high symmetry of QD confining potential is important for efficient generation of entangled photon pairs, as it directly impacts the bright-exciton fine structure splitting (FSS)~\cite{Senellart2017}.
Symmetry can be inherited from the substrate by employing high-Miller-index planes, such as $(111)$A-~\cite{Kuroda2013} and $(111)$B-oriented GaAs~\cite{Juska2013} for the emission in the visible range, or InP$(111)$A with C$_{3v}$ symmetry~\cite{Liu2014,Ha2016} in the near-infrared spectral range.
These high-Miller-index planes substrates are difficult to process and are thus unpractical for device applications.
Moreover, despite the promising prospects of this approach~\cite{Schliwa2009}, the FSS of the resulting QDs have not been below ${(60\pm 38)}~\si{\micro\electronvolt}$~\cite{Jahromi2021}.
Therefore, it is important to develop new methods that allow synthesis of symmetric QDs on the industry-compatible $(001)$-oriented InP substrate.
Inspiring results have been achieved in the GaAs-based material system using the DE approach~\cite{Wang2006,Wang2007}.
The obtained DE QDs have been shown to possess high in-plane symmetry and hence, exhibit very low FSS values~\cite{Huo2013,Heyn2009}, allowing them to be employed in quantum communication schemes utilizing entangled photon pairs~\cite{Basset2021}. For $(001)$ InP substrates, a modified DE method, where the QD crystallization from an In droplet is followed by the annealing in AsH$_3$ ambient, was recently suggested by Sala \textit{et al.}~\cite{Sala2020}. 
The resulting QDs were shown to have high in-plane symmetry and are located at the center of etched pits, formed during the annealing step.

In this work, we present a detailed investigation of the morphology, chemical composition, electronic structure and optical properties of ensemble, and individual droplet epitaxy InAs(P)/InP QDs grown in metalorganic vapour-phase epitaxy (MOVPE). By optimizing the growth process, we could achieve an array of low surface density QDs ($\SI{2.8e8}{\centi\meter\tothe{-2}}$), with highly in-plane symmetric profiles located in etched pits. The QD emission covers the range of interest centered around $\SI{1500}{\nano\meter}$. We have examined in detail the morphology and chemical composition of surface and buried QD-in-pit structures using atomic force microscopy (AFM) imaging, scanning transmission microscopy (STEM) and energy dispersive X-ray spectroscopy (EDX). Based on the insight gained, we propose a kinetic model to explain the formation of asymmetric pits etched around the QDs and the QD base shape modification. Moreover, we have performed theoretical calculation of electronic and optical properties within the eight-band $\kp$ method combined with the configuration-interaction approach for states of carrier complexes. Together with the morphological investigations they support the interpretation of properties of ensemble and individual QDs revealed by the low-temperature photoluminescence (PL) measurements. Excitonic complexes in individual QDs were studied by the micro-PL ($\upmu$PL) and time-resolved $\upmu$PL experiments. We show that the fabricated QDs are promising candidates for single-photon sources emitting at $\SI{1500}{\nano\meter}$, supported by their high purity single-photon emission with an as-measured value of $\mathcal{P}=92.5\%$ and fitted $g^{(2)}_{\mathrm{fit}}=0$ with a standard error of $\sigma=0.10$ associated with the fitting procedure. In addition, the high lateral symmetry of the QDs makes them interesting for entangled photon pair generation for quantum communication applications. Although we measure an exciton FSS value on the level of $(50\pm5)~\si{\micro\electronvolt}$, this is likely related to the post-growth processing of the QD structure and require further optimization.

\section{Results and discussion}
\subsection{Description of the structure}
Each of two studied samples contains two QD arrays grown under the same conditions by low-pressure MOVPE on an $(001)$-oriented InP substrate.
The first QD array is covered by a $\SI{30}{\nano\meter}$-thick InP layer and was used for STEM and optical studies.
The second QD array is deposited on the top InP surface and is left uncovered for AFM investigation.

The QDs were synthesized in a two-step process. 
First, an array of In droplets was deposited on the InP surface at $\SI{360}{\degreeCelsius}$ under a trimethylindium (TMIn) flux and in the absence of V\textsuperscript{th} group flux.
Afterwards, the In droplets were annealed under an arsine (AsH$_3$) ambient as the temperature was being ramped up to $\SI{550}{\degreeCelsius}$, followed by $\SI{180}{\second}$-long waiting time (annealing).
For comparison, a reference structure was grown without the high-temperature annealing.
Instead, after the In droplet deposition at $\SI{360}{\degreeCelsius}$, the temperature was raised only to $\SI{475}{\degreeCelsius}$ in AsH$_3$ ambient and no waiting followed.
A detailed description of the growth procedure can be found in the Methods section.

\begin{figure*}
 \centering
 \includegraphics[width=0.75\linewidth]{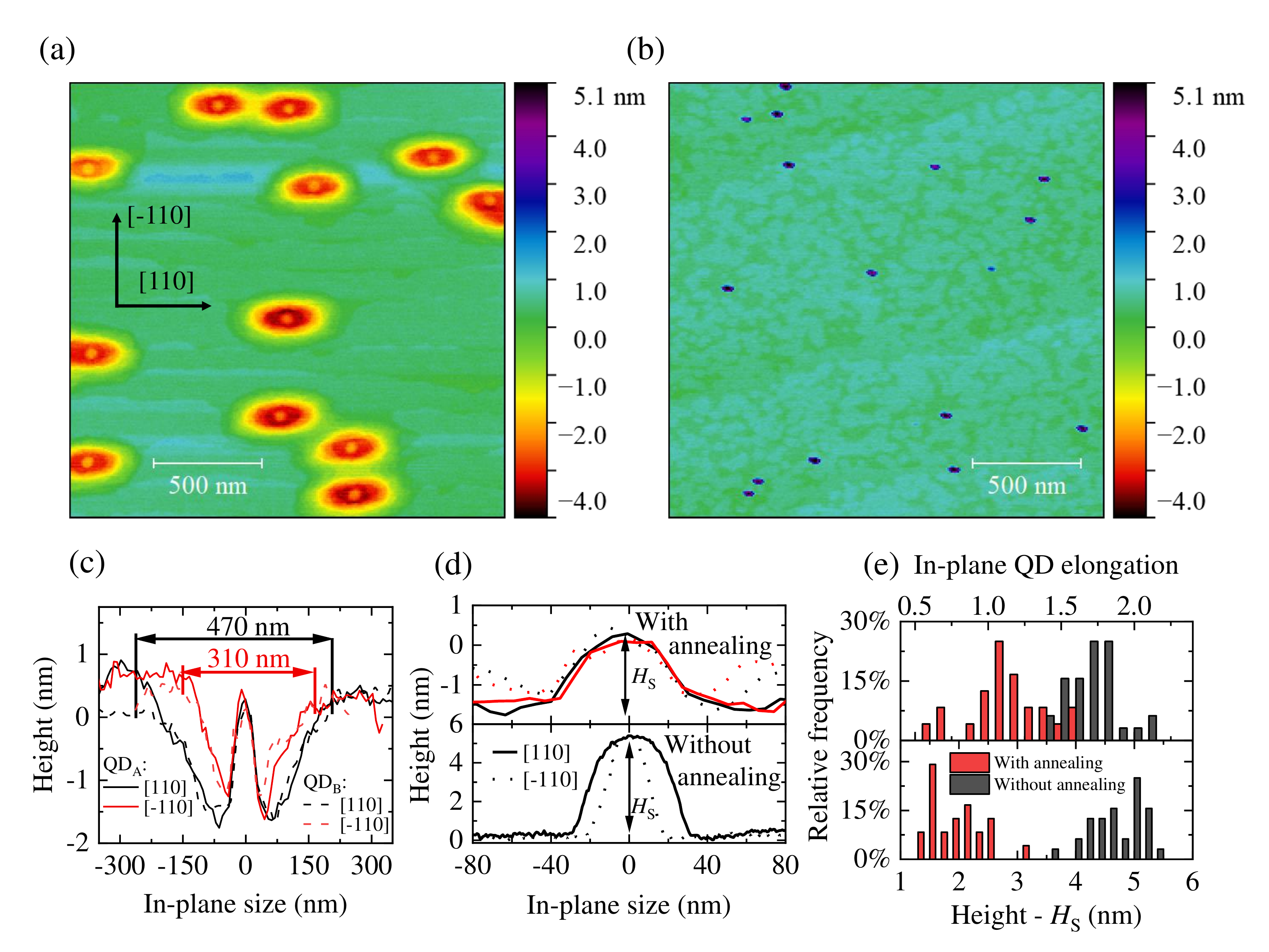}
 \caption{Morphological investigation of surface DE QDs by atomic force microscopy, ${(2\times2)}~\si{\micro\meter\tothe{2}}$ scans.
 (a) Annealed QDs in etched pits.
 (b) QDs before annealing.
 (c) Cross-section scans of the pits around annealed QDs.
 (d) Comparison of the QD cross-sections: after annealing [top panel, close-up of (c)] and before annealing (bottom).
 (e) Statistics of the QD in-plane elongation (top) and height (bottom).}
 \label{fig:AFM}
\end{figure*}

\subsection{QD morphology investigations}
The surface morphology of the QDs was investigated by AFM.
\subfigsrefL{AFM}{a}{b} present typical AFM images of the surface QDs in the annealed, and the un-annealed reference samples, respectively.
The number density of the surface QDs is similar in the two samples, since this parameter is determined by the initial stage of DE QD growth, i.e. the In droplet deposition~\cite{Gurioli2019}.
The difference between the two samples arises from the annealing conditions under AsH$_3$.
It can be seen that for the sample annealed at an elevated temperature, every QD is located in a pit formed by local etching of the surrounding material on the perimeter of the InAs DE QD~\cite{Sala2020}.
The exemplary cross-sectional profiles of the AFM scans through the QD centers for the annealed sample shown in \subfigrefL{AFM}{c} reveal the typical shape of the pits in two orthogonal directions.
The pits are asymmetric, with a typical width of $\SI{\sim470}{\nano\meter}$ and $\SI{\sim310}{\nano\meter}$ along the [110] and $[\overline{1}10]$ directions, respectively.
The high-resolution QD AFM profiles in \subfigrefL{AFM}{d} reveals the high symmetry of the annealed QDs (top panel) in comparison to the reference QDs (bottom panel).
Statistical analysis of the QD profiles presented in \subfigref{AFM}{e} shows that the in-plane aspect ratio of the annealed QDs, defined as the ratio of the QD base median length along $[110]$ and $[\overline{1}10]$ directions, is $1.09$ compared to $1.75$ for the reference QDs.
We also observe a significant decrease of the QD height due to annealing, from $H_s\approx\SI{4.8}{\nano\meter}$ for the reference sample to $H_s\approx\SI{1.9}{\nano\meter}$ for the annealed QDs [bottom panel in \subfigref{AFM}{e}].
For the annealed QDs, we defined $H_s$ as the height of a surface QD measured from the bottom of the pit to the QD apex.

\begin{figure*}
 \centering
 \includegraphics[width=0.7\linewidth]{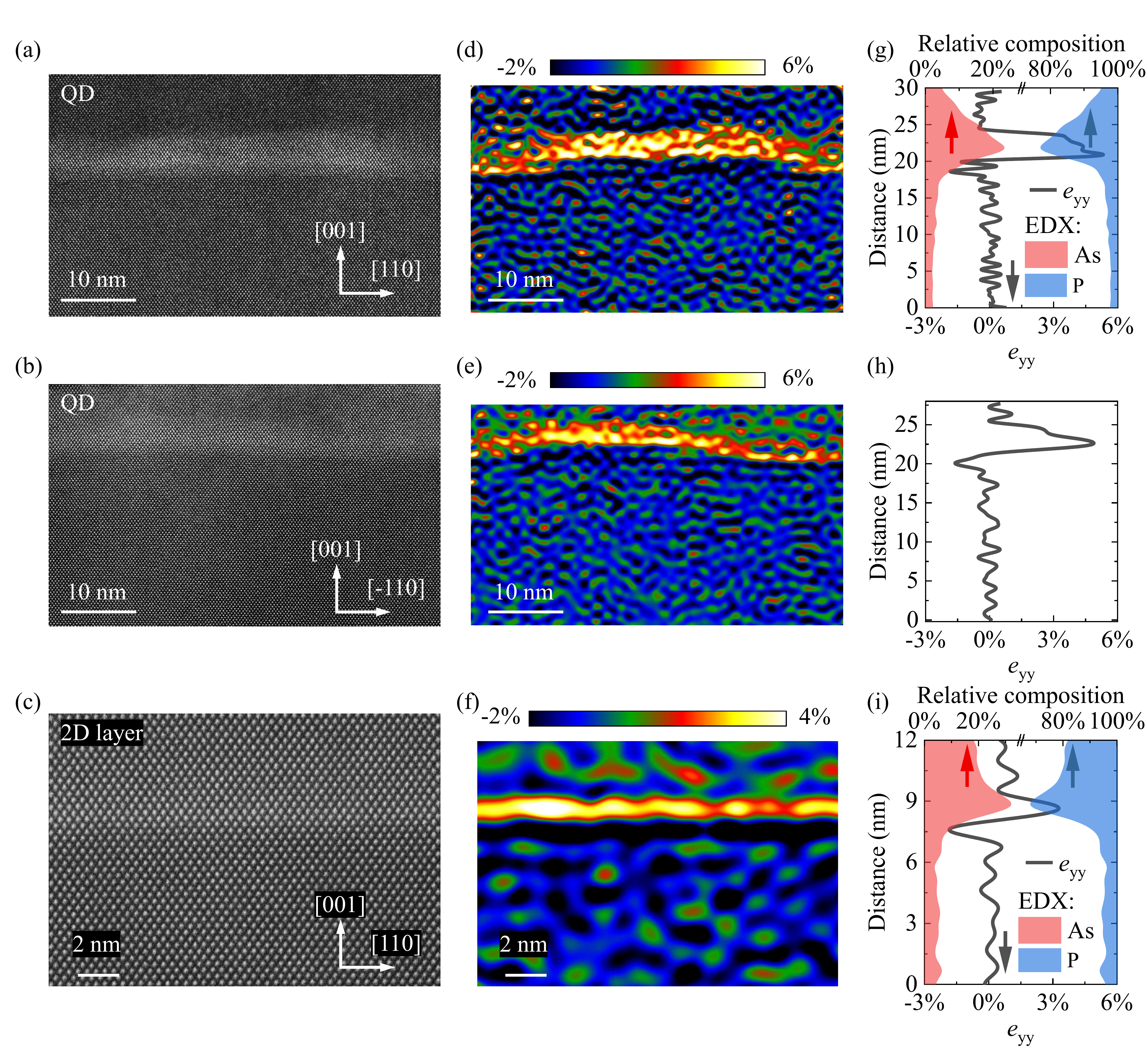}
 \caption{Morphological investigation of the buried annealed DE QDs by STEM and EDX.
 QDs in cross-section geometry viewed along the (a) $[\overline{1}10]$ and (b) $[110]$ directions, and (c) the 2D InAs(P) layer.
 (d)-(f) Maps representing the fractional change in lattice spacing along the $[001]$ direction (relative to the InP lattice), $e_{yy}$, for each of the images in (a)-(c).
 (g)-(i) Line scan profiles of P (blue) and As (red) concentrations measured by EDX, and the lattice displacement values $e_{yy}$ (black lines) along the $[001]$ direction.
 }
 \label{fig:TEM}
\end{figure*}

Since surface QDs are optically inactive, buried QDs need to be examined with respect to photon generation processes.
Detailed investigation of the shape and composition of annealed buried QDs and their environment was carried out here by high resolution STEM~\cite{Kadkhodazadeh2013} and EDX in cross-section geometry.
The results are summarized in \figref{TEM}, where we show cross-sections of QDs viewed along the $[\overline{1}10]$ (top row) and $[110]$ directions (middle row), and on the 2D layer (bottom row).
While the base length of the buried QDs seems to be very similar to the surface QDs, we measure larger heights for the buried QDs compared to the surface ones.
Based on the images, the buried QDs can be approximated as truncated cones with arched bases and almost symmetric in-plane dimensions.
We measure a base diameter of $B = {(45\pm2)}~\si{\nano\meter}$, top diameter of $D ={(33\pm1.7)}~\si{\nano\meter}$ and height of $H = {(5.2\pm0.3)}~\si{\nano\meter}$ for the buried QDs in the annealed sample.
The chemical composition of these QDs is found to be InAs$_{x}$P$_{1-x}$ with $x = (80 \pm 15)\%$, measured both by EDX and by analyzing the change in the lattice parameter along the $[001]$ direction from the atomic resolution STEM images (for details, see Methods).
The color-coded maps of changes in lattice spacing along the $[001]$ direction relative to the InP substrate ($e_{yy}$) are shown in \subfigsref{TEM}{d}{f}.
Line scans of $e_{yy}$ across the QDs and As and P composition profiles obtained from EDX measurements are plotted in \subfigsref{TEM}{g}{i}.
Additional EDX data is shown in \Supplfigref{EDX-QD}.

\begin{figure}[t]
 \centering
 \includegraphics[width=\linewidth]{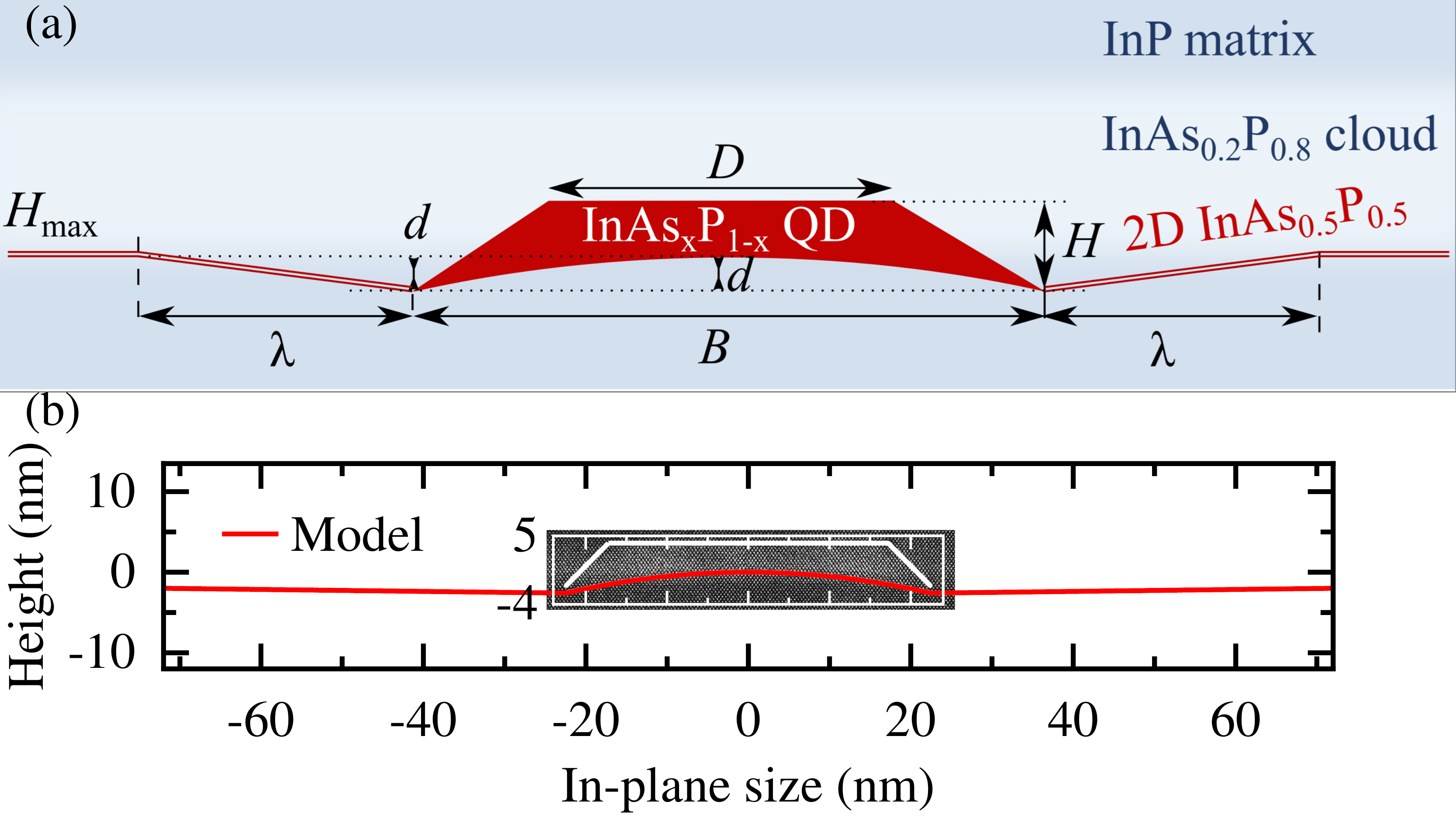}
 \caption{(a) Model of a QD and surrounding pit inferred from the STEM scans.
 (b) The modelled profile of the bottom base of the QD and the 2D InP(As) layer surrounding it (red line) overlaid on the STEM image in \subfigref{TEM}{a}.
 The white eye-guiding line indicates the QD top interfaces.}
 \label{fig:QDscheme}
\end{figure}

In contrast to SK QDs, DE QDs are formed without a wetting layer, due to the entirely different nature of their nucleation~\cite{Skiba-Szymanska2017}.
In the case of InAs/GaAs QDs, the dot and the barrier share the V\textsuperscript{th} group atoms, while in the case of our studied InAs/InP material system, the group V\textsuperscript{th} fluxes are switched at the QD interface.
This means that the InP surface is exposed to the AsH$_3$ ambient during the QD crystallization step.
As the V\textsuperscript{th} group atoms are volatile, P atoms get desorbed from the surface and are substituted by As atoms, thus forming a 2D layer of InAs$_{x}$P$_{1-x}$~\cite{Gonzalez2002,Carlsson1998}.
The thickness of this layer in our sample is $h \approx 3$~MLs, estimated from the STEM images [see \subfigref{TEM}{c}, \subfigref{TEM}{f}, and \subfigref{TEM}{i}].
A chemical composition of $x = (50 \pm 15)\%$ is estimated for this layer from both atomic resolution STEM images and EDX profiles as shown in \subfigref{TEM}{f} and \subfigref{TEM}{i}. 
For EDX data of the 2D layer, see \Supplfigref{EDX-2D-layer}.
Moreover, we consistently observe a region about $\SI{15}{\nano\meter}$-thick above the 2D layer containing up to $\sim20\%$ As.
Based on our AFM and STEM observations of DE QDs and their environment, we reconstruct the model presented in \subfigref{QDscheme}{a} for buried DE InAs/InP QDs located in etched pits and connected by a 2D InP(As) layer.

In the following section, we focus on an analytical description of the local etching mechanism during the DE QD annealing step.
We assume a symmetric QD with dimensions inferred from the STEM measurements (QD1).
We assume the bottom base diameter of $B\sim\SI{43}{\nano\meter}$ and top diameter of $D\sim\SI{33}{\nano\meter}$.
We then include a small QD elongation of $\sim1.09$ to the model, as measured by AFM for surface QDs.
This resizes the dimensions to $B_{\mathrm{[\overline{1}10]}}\sim\SI{41.1}{\nano\meter}$ and $D_{\mathrm{[\overline{1}10]}}\sim\SI{31.6}{\nano\meter}$
along the $[\overline{1}10]$ direction, and
$B_{\mathrm{[110]}}\sim\SI{44.9}{\nano\meter}$
and $D_{\mathrm{[110]}}\sim\SI{34.4}{\nano\meter}$ along the $[110]$ direction (QD2).

Interestingly, STEM images show that the bottom base of the QD is arched upwards, where the sides are at the same level as the etched pits around the QDs and the central part of QD base is approximately at the same level as the 2D layer outside the pits.
Therefore, in our model, we keep the pit’s depth and the central part of the QD base level equal [marked with $d$ in \subfigref{QDscheme}{a}].
For QD1, we take $d=\SI{1.9}{\nano\meter}$ in both directions, and for QD2 $d=\SI{2.6}{\nano\meter}$ and $d=\SI{2}{\nano\meter}$ along the $[110]$ and $[\overline{1}10]$ directions, respectively.
For both QDs, we keep the same in-plane distance $\lambda$ between the QD base and the substrate outside the pit: $\lambda = \SI{213}{\nano\meter}$ along $[110]$ and $\lambda = \SI{133}{\nano\meter}$ along the $[\overline{1}10]$ direction. 

\subsection{Formation of pits around QDs}
Annealing takes place under the As supplying ambient, while P is available from the InP substrate.
We need, however, an In supply to form additional InAs(P) material in the island during annealing, most probably followed by InAs(P) evaporation from the QD apex, leading to the observed decrease in the height of the annealed QDs. The material exchange between the substrate and the island is In-limited. We assume that the In diffusion occurs preferentially through the perimeter of the island base in contact with a 2D InP(As) layer, where the elastic stress induced by the lattice mismatch is highest, similar to InAs/InGaAs system~\cite{Osipov2002,Dubrovskii2003}. Coordinate-dependent coverage of the substrate with In atoms $\theta$ can be described using the steady-state diffusion equation
\begin{equation}
    \frac{d^2\theta}{dx^2}=-I_k\Omega/D_k h,
\end{equation}
with $k$=0 or $1$. Here, $x$ is the coordinate along a given crystallographic direction (either $[\overline{1}10]$ or $[110]$), $\Omega$ is the elementary volume per one III-V pair, $h$ is the height of InAs(P) ML, $I_0$ and $D_0$ are the In fluxes [$\si{\nano\meter\tothe{-2}\second\tothe{-1}}$] and diffusion coefficient for In migration from the substrate to area around the island, while $I_1$ and $D_1$ describe In diffusion from the substrate into the island through the interface beneath the QD.
The diffusion process is driven by the elastic stress field~\cite{Osipov2002,Dubrovskii2003}, which depends on the position around the island and the shape of the QD itself.

The relevant boundary conditions are given by 
\begin{equation}\label{eq:eq2}
    \theta(x=B/2)=\theta_{\mathrm{min}},\quad \theta(x=B/2+\lambda)=\theta_{\mathrm{max}} 
\end{equation}
for $B/2\leq x\leq B/2+\lambda$ and
\begin{equation}\label{eq:eq3}
    \left(\frac{d\theta}{dx}\right)_{x=0}=0,\quad \theta(x=B/2)=\theta_{\mathrm{min}}
\end{equation}
for $0\leq x \leq B/2$. Here, $B$ is a QD base diameter along a given direction and $\lambda$ is the effective diffusion length of In atoms in this direction.  The height profile of the surface around a QD is then obtained as 
$H-H_{\mathrm{max}}=-I_k\Omega t (\theta_{\mathrm{max}}-\theta)$, with $H_{\mathrm{max}}$ as the height outside the pit (the zero level) and $t$ as the annealing time.
According to \eqnref{eq2}, the profile depth is maximum at the island boundary  and zero outside the pit.
According to \eqnref{eq3}, the height of the surface layer in the growth direction reaches its maximum at $x=0$ by symmetry.
Our measurements show that this maximum is very close to the surface height outside the pit, meaning that the etching does not occur under the QD, e. g., in the center of the pit. 
Solutions for the pit cross-sections along the $[\overline{1}10]$ and $[110]$ directions are given by
\begin{multline}\label{eq:eq4}
    H-H_{\mathrm{max}}=
    -I_0\Omega \Delta\theta t \left[ 1-\frac{x-B/2}{\lambda}\right.\\
    \left.-\frac{(B+\lambda)(x-B/2)+B^2/4-x^2}{r_0^2}\right]
\end{multline}
for $B/2\leq x \leq B/2+\lambda$ around the QD and

\begin{equation}
    H-H_{\mathrm{max}}=-I_1\Omega \Delta\theta t \left[ 1-\frac{B^2/4-x^2}{r_1^2}\right]
\end{equation}
for $0 \leq x \leq B/2$ beneath the QD, with 
$r_i^2=2D_i h\Delta\theta/(I_i \Omega)$ ($i=0,1$), and $\Delta\theta=\theta_{\mathrm{max}}-\theta_{\mathrm{min}}$.
According to the measurements, the height profiles of the pits around the QD are linear within the AFM experimental accuracy (the slope $d/\lambda$ is only $1-2\%$), corresponding to $r_0\to\infty$ in \eqnref{eq4}.
This should be due to a high diffusivity of In through 2D layer relative to the diffusivity of In at the island-substrate interface ($D_0\ll D_1$).
The calculated profile in \Supplfigref{Nucleation-theory} shows the excellent fits to the measured cross-sections of the pits around symmetric QD1 and asymmetric QD2, as well as the concave-shaped cross-sections of the QD base.
The fits are obtained with $I_0\Omega \Delta\theta t=I_1 \Omega \Delta\theta t = \SI{1.9}{\nano\meter}$, $B=\SI{43}{\nano\meter}$, $r_1=\SI{21.5}{\nano\meter}$,
$\lambda=\SI{132.5}{\nano\meter}$ in the $[\overline{1}10]$ direction and $\SI{213}{\nano\meter}$ in the $[110]$ direction for symmetric QD1.
For asymmetric QD2, we use the same $\lambda$, $I_0\Omega \Delta\theta t=I_1\Omega \Delta\theta t=\SI{2.0}{\nano\meter}$, $B=\SI{41.12}{\nano\meter}$, $r_1=\SI{20.58}{\nano\meter}$ in the $[\overline{1}10]$ direction, and $I_0\Omega \Delta\theta t=I_1\Omega \Delta\theta t=\SI{2.6}{\nano\meter}$, $B=\SI{44.88}{\nano\meter}$, $r_1=\SI{22.4}{\nano\meter}$ in the [110] direction.
In all cases, the values of $r_1$ are very close to $B/2$, which explains an absence of etching in the center of the pit.
The fitting values of $I_0\Omega \Delta\theta t$ and $I_1\Omega \Delta\theta t$ appear identical, showing that the supply of In from the substrate to the QD volume is the same around and underneath the QDs, while different shapes of the cross sections are due to different diffusion coefficients of In atoms.
The effective diffusion length of In atoms is larger along the $[110]$ direction thus the etching process occurs at different rates along different crystallographic directions and results in the elongated pit profiles.
Overall, the fact that QDs become almost symmetric after annealing should be due to a more homogeneous solid diffusion of In atoms compared to its surface diffusion leading to elongated shapes of the initial islands. 

\subsection{Electronic and optical properties of QDs}

\subsubsection{Optical properties of the QD ensemble} 

\begin{figure}[t]
 \centering
 \includegraphics[width=\linewidth]{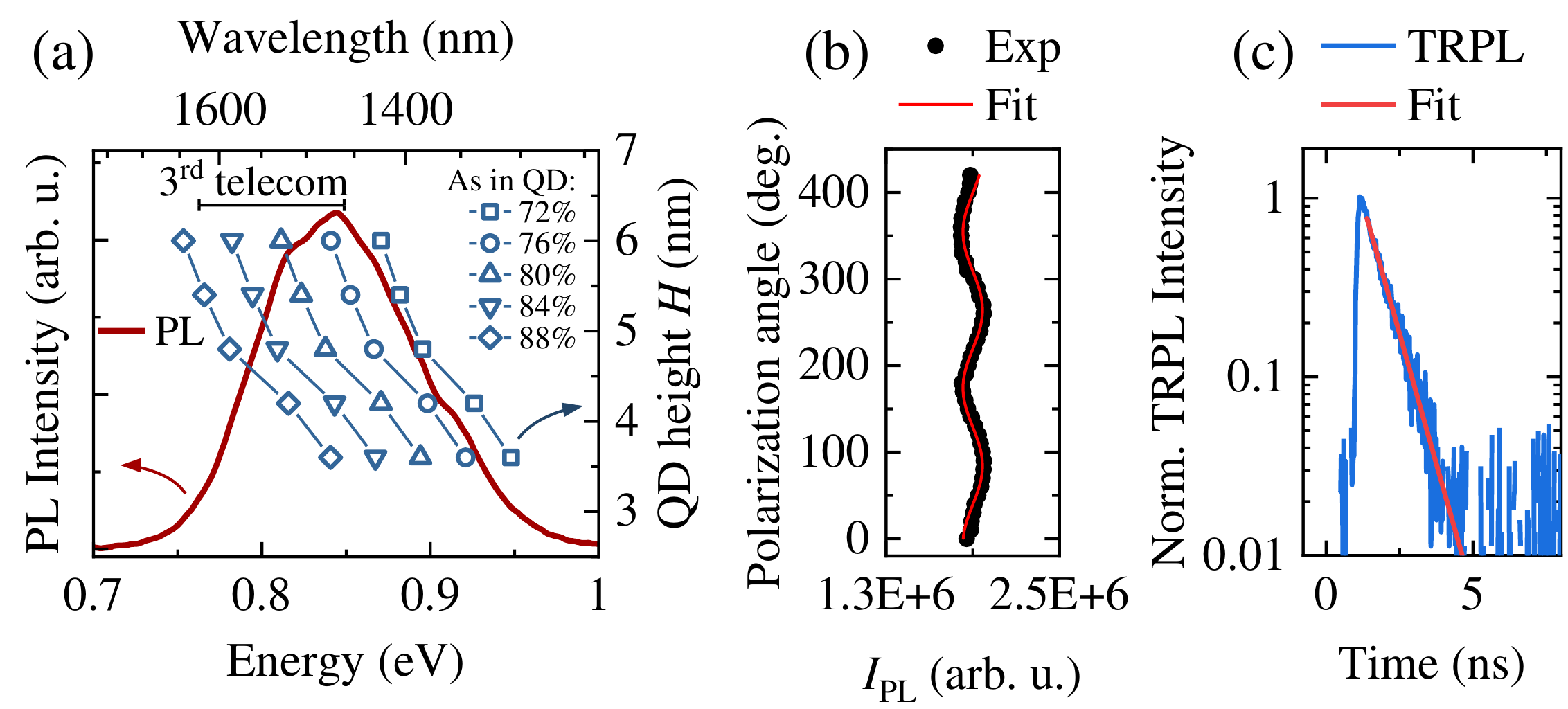}
 \caption{Optical properties of the QD ensemble emission.
 (a)~PL spectrum (left axis) with the calculated X emission energy as a function of QD height and As content (right axis).
 (b)~Time-resolved photoluminescence for the peak of the QD ensemble.
 (c)~Integrated PL intensity of QD ensemble as a function of the polarization angle.}
 \label{fig:Optical-ensemble}
\end{figure}

Optical properties of an ensemble of buried QDs were investigated by low-temperature ($T=\SI{12}{\kelvin}$) PL, time-resolved PL and polarization dependence of QD PL response, and the results are summarized in \subfigref{Optical-ensemble}.
The optical response of the QD ensemble reflects the statistical properties of the investigated QDs.
The PL spectrum of the sample is shown in \subfigref{Optical-ensemble}{a}.
The emission is centred at $\SI{\sim0.85}{\electronvolt}$, showing spectral broadening mainly related to the QD size distribution and chemical composition. 

To link the optical and morphological investigations of the QDs, we calculate the electron and hole eigenstates within the multiband envelope-function $\kp$ theory, based on the QD geometry presented in \subfigref{QDscheme}{a}, and then include Coulomb and anisotropic electron-hole exchange interactions within the configuration-interaction approach.
Details of the numerical calculations and the parameters used are given in the Methods section.
We probe a range of parameters within the uncertainty window provided by the morphological studies of buried annealed QDs.
The exciton ground-state energy for a series of calculations for QDs with As composition $x$ in the range $\SIrange{72}{88}{\percent}$ and height $H$ varied between $\SI{3.6}{\nano\meter}$ and $\SI{6.0}{\nano\meter}$, is plotted with symbols in \subfigref{Optical-ensemble}{a} for comparison with the PL spectrum.
The position of the PL peak fits to an exciton transition calculated for an InAs$_{x}$P$_{(1-x)}$ QD with $x\sim\SI{80}{\percent}$ and $H \sim \SI{5}{\nano\meter}$, thus confirming a good agreement between the QD model constructed from the HR STEM and EDX measurements and the optical response.

The degree of linear polarization (DOLP) for the ensemble emission, presented in \subfigref{Optical-ensemble}{b}, is experimentally estimated to be on the low level of $\sim3.5\%$ (see Methods for the definition of DOLP).
Such a low value can be linked to a very small in-plane asymmetry of the confining potential.
This parameter is influenced by strain fields, local atomic disorder, the electric field, and more importantly, the QD shape~\cite{Gawelczyk2017}.
High in-plane symmetry of QDs revealed from the morphological investigation is expected to provide a highly symmetric QD potential and, consequently, lead to vanishing fine structure splitting for confined exciton and biexciton states.
This satisfies the general requirement for a source for polarization-entangled photon pairs.

We also carried out time-resolved PL measurements of QD ensemble to estimate the recombination time at the PL peak energy.
The obtained PL trace shows a mono-exponential decay characterized by a time constant $\tau_{\textrm{PL}}=(0.75\pm0.01)~\si{\nano\second}$ (\subfigref{Optical-ensemble}{c}).
In the case of QD confining potential asymmetry, two PL decay components in the PL trace should be observed due to two different oscillator strengths for the doublet of fundamental exciton transitions~\cite{Gawelczyk2017}.
This in turn would result in a significant DOLP.
Thus, the results of optical investigations of the QD ensemble are in line with the expected relatively small fine structure splitting in the investigated QDs.

\subsubsection{Single QD emission properties}

\begin{figure}[t]
 \centering
 \includegraphics[width=\linewidth]{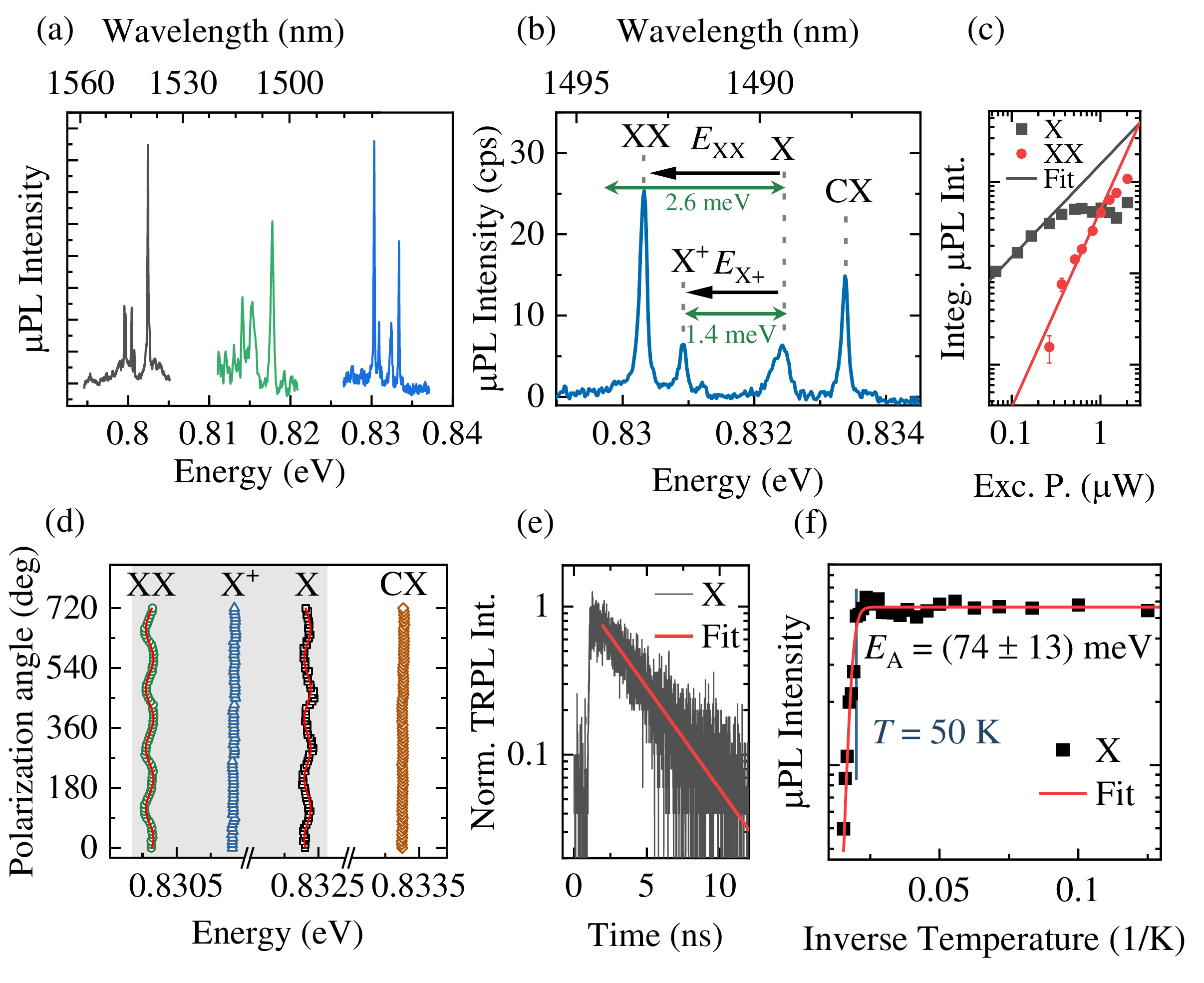}
 \caption{Single QD emission.
 (a)~Exemplary high-resolution low-temperature ($T = \SI{4.2}{\kelvin}$) $\upmu$PL spectra measured in three different mesas containing QDs.
 (b)~$\upmu$PL spectrum of an individual QD with identified excitonic complexes (exciton, X, positive trion, X$^{+}$, and biexciton, XX), as well as a trion line from a neighboring QD (CX).
 The corresponding calculated binding energies for trion and biexciton are written and shown with green arrows (see QD2 in \tabref{calc}).
 (c)~Integrated $\upmu$PL intensity for XX and X lines with fit lines.
 (d)~Polarization-resolved $\upmu$PL for the lines shown in panel (b).
 (e)~Time-resolved photoluminescence for the X line.
 (f)~Temperature dependence of the X line $\upmu$PL intensity with the Arrhenius fit line.}
 \label{fig:Optical1}
\end{figure}

\begin{table*}[t]
\small
  \caption{Results of calculation for the idealized (QD1) and realistic (QD2) models of a QD: first excited level splitting for the electron and the hole, light-hole admixture, X ground state energy, X lifetime, carrier-complex binding energies}
  \label{tab:calc}
  \begin{tabular*}{\textwidth}{@{\extracolsep{\fill}}ccccccccccc}
    \hline
    & $e_2-e_1$ ($\si{\milli\electronvolt}$) & $h_2-h_1$ ($\si{\milli\electronvolt}$) & lh admixture (\%) & $E_{\mathrm{X}}$ ($\si{\milli\electronvolt}$) & DOLP (\%) & $\tau_{\mathrm{X}} (\si{\nano\second})$ & $\Delta_{\mathrm{X}^{-}}$  ($\si{\milli\electronvolt}$) & $\Delta_{\mathrm{X}^{+}}$ ($\si{\milli\electronvolt}$) & $\Delta_{\mathrm{XX}}$ ($\si{\milli\electronvolt}$) \\
    \hline
    QD1 & 10.18 &	7.64&	0.056&	823.82 & 0.32 &	0.87	&3.01	&1.64	&2.78\\
    QD2	&7.60	&7.77	&0.06	&818.75 & 1.42	&0.84	&2.99	&1.43 &2.63\\
    \hline
  \end{tabular*}
\end{table*}

In this section, we present a detailed investigation of the optical properties of individual QDs.
To spatially isolate individual QDs, the structure was processed into large mesas $\SIrange{2}{3}{\micro\meter}$ in size (for details see Methods).
The $\upmu$PL spectra for the buried annealed QDs are presented in \subfigref{Optical1}{a}, comprising of a handful of well-isolated emission lines, indicating the presence of only a few QDs within the excitation spot.

A zoomed-in plot of the group of $\upmu$PL lines at $\sim\SI{0.83}{\electronvolt}$ is presented in \subfigref{Optical1}{b}.
The excitonic complexes are identified by measuring the excitation power ($P$) dependence of the line intensity ($I$). While for the $\upmu$PL line at $\sim\SI{0.8324}{\electronvolt}$ we have $I_{\mathrm{X}}\sim P^{1.01}$, for the line at $\sim\SI{0.8303}{\electronvolt}$ the intensity changes quadratically with power ($I_{\mathrm{XX}}\sim P^{1.95}$). 
These dependencies, displayed in \subfigref{Optical1}{c}, allowed for a tentative assignment of the lines as exciton (X) and biexciton (XX) transitions in the same QD. 
The $\upmu$PL spectrum also contains another line at $\sim\SI{0.8334}{\electronvolt}$, for which $I_{\mathrm{CX}}\sim P^{1.27}$.
Therefore, the line is tentatively identified as a charged exciton transition (CX)~\cite{Abbarchi2009,Baier2006}, without clear assignment to a particular QD. 

We carry out theoretical calculations of a QD having the ground state near the observed X transition.
The geometry of simulated QDs is based on our morphological investigations (see  \figref{QDscheme}) with $H=\SI{4.8}{\nano\meter}$ and $x=\SI{80}{\percent}$, while for the in-plane size we calculate both an ideally symmetric (labeled QD1) and a slightly asymmetric one (QD2), which includes minor deviations from the perfect symmetry: slight elongation and a difference in the concave QD base arcs in the $[110]$ and $[\overline{1}10]$ directions.
All the relevant computed parameters are summarized in \tabref{calc}.
The only parameter with a noticeable, yet still insignificant, difference between the ideally symmetric QD1 and the realistic QD2 is the electron single-particle excitation energy.
Thus, we can consider that the slight asymmetry present in the investigated QDs should have very small effects on optical properties.

The calculated binding energies for the biexciton (XX), the positively-charged exciton (X$^{+}$), and the negatively-charged exciton (X$^{-}$) are negative with magnitudes of $\Delta_{\mathrm{XX}}=\SI{2.63}{\milli\electronvolt}$, $\Delta_{\mathrm{X}^{+}}=\SI{1.43}{\milli\electronvolt}$, and $\Delta_{\mathrm{X}^{-}}=\SI{2.9}{\milli\electronvolt}$, respectively (see \Supplfigref{calc-binding-en} for the binding energies for all calculated QD2 geometries).
The calculated $\Delta_{\mathrm{XX}}$ value fits very well to the experimentally obtained XX binding energy, whereas another line in the QD spectrum fits to the X$^{+}$ binding energy.
For comparison, the calculated binding energies are shown in \subfigref{Optical1}{b} with green arrows.
Therefore, the respective exciton complexes presumably belong to the same QD. 
However, the CX line at emission energy above X is attributed to a different QD as the existence of a negative trion with a positive binding energy may be excluded based on the theoretical calculations.

The results of polarization-resolved $\upmu$PL investigation are presented in \subfigref{Optical1}{d}.
Since the identified X$^{+}$ and CX lines come from recombination of spin-singlet states, one can expect zero FSS, and hence no doublet in the linear polarization-resolved spectrum.
However, such doublets are present for the X and XX emission lines. The extracted exciton FSS is ${\sim(50\pm5)}~\si{\micro\electronvolt}$, which is comparable to the average values in symmetric GaAs-based QDs~\cite{Jahromi2021}.
However, it is much higher than expected for the studied DE QDs as suggested by the ensemble emission properties described above. 
Moreover, the minor deviation from rotational symmetry found in the QD morphology, according the calculation summarized in \tabref{calc}, could not explain the appearance of considerable FSS, since it results in negligible light-hole admixture to the hole ground state (see \tabref{calc}), which is the main source of FSS in QDs~\cite{Tsitsishvili2017}.
Additionally, time-resolved $\upmu$PL investigations of the exciton line, shown in \subfigref{Optical1}{e}, reveal $\tau=(3.16\pm0.04)~\si{\nano\second}$. This is in stark contrast to the ensemble measurements, $(0.75\pm0.01)~\si{\nano\second}$, and to the theoretically predicted value of $0.84~\si{\nano\second}$, which indicates the presence of other factors involved in carrier states and dynamics in the processed sample.
The measured FSS and increased QD lifetime can be attributed to, e. g., the presence of crystal point defects and/or electric charges in the vicinity of QDs as a consequence of the dry etching process for mesa fabrication. This effective degradation of the initial in-plane rotational symmetry of the QD confining potential (exhibited in FSS and recombination time) requires further investigation.

The temperature–dependent $\upmu$PL intensity of the X line is used to evaluate the prospect of single-photon applications at elevated temperatures, which require only Stirling-compatible cryocooler~\cite{Musial2020} (\subfigref{Optical1}{d}).
Quenching of the $\upmu$PL intensity is observable at relatively high temperatures as for pure InP barriers, starting from $T=\SI{50}{\kelvin}$.
The activation energy, $E_A$, of ${(74\pm13)}~\si{\milli\electronvolt}$ (see Methods for the fitting formula), is larger than typically observed for SK InAs/InP QDs~\cite{Holewa2020PRB} or InAs/InAlGaAs quantum dashes~\cite{Dusanowski2016}, and proves a good carrier localization in the QDs under investigation.

\begin{figure}[t]
 \centering
 \includegraphics[width=\linewidth]{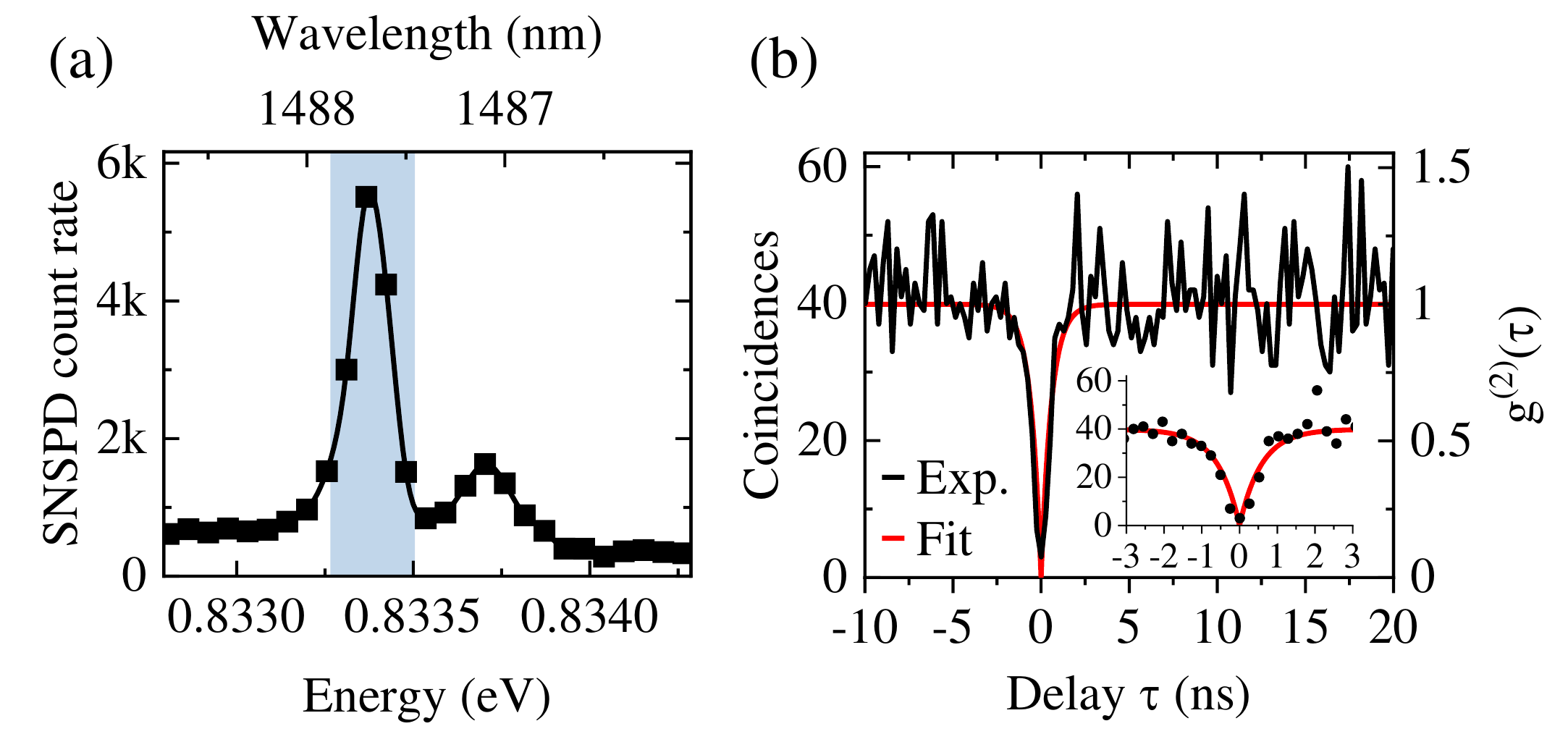}
 \caption{(a)~The $\upmu$PL spectrum for the CX line recorded on superconducting nanowire single-photon detectors with (b)~CX autocorrelation histogram.
 Shaded area in (a) marks the monochromator band-pass.}
 \label{fig:Optical2}
\end{figure}

Finally, we investigate the single-photon emission purity $\mathcal{P}$ by the autocorrelation spectroscopy of the charged-exciton line in \subfigref{Optical2}{a}, having one of the largest emission intensities. 
For that, we record the second-order correlation function $g^{(2)}(\tau)$, exploiting the off-resonant cw excitation scheme.
The obtained histogram $C(\tau)$ without normalization is presented in \subfigref{Optical2}{b} which we fit~\cite{Michler2000a} (see Methods) and obtain the $g^{(2)}(\tau)$ function by normalizing $C(\tau)$ with the average level $A$ of coincidences per channel for $|\tau|\gg0$.
We obtain the fitted value of $g^{(2)}(\tau=0)=0$ and the standard error of the fitting procedure $\sigma=0.10$ (without background correction or other data post-processing and with $A=40$).
The raw data-estimated purity is $\mathcal{P}=1-C(0)/A=\SI{92.5}{\percent}$, and with the moderate level of $A$, we employ a conservative estimation of the single-photon purity $\mathcal{P}=(92.5 \pm 7.5)\%$.
The obtained high purity of the single-photon emission indicates the potential of the investigated DE QDs in quantum information processing as single-photon emitters operating at the third telecom window.

\section{Conclusions}
We have synthesized InAs/InP QDs with low-surface density on $(001)$-oriented InP substrate by droplet epitaxy in the MOVPE process.
By implementing an additional annealing step under AsH$_3$ ambient after QD formation, we induced modifications to the QD shape, as well as local etching of pits around the QD's perimeter.

The morphology of resulting DE QDs was investigated in detail by atomic force microscopy, scanning transmission electron microscopy and energy dispersive X-ray spectroscopy.
The resulting DE QDs were found to have close to symmetric in-plane profiles and concave bases. 
Based on our morphological investigations, we proposed a kinetic model describing the formation of the pits surrounding QDs and the modification of the QD base shape.

The optical properties of a DE QDs ensemble reveled low degree of linear polarization, supporting the observation of high in-plane symmetry of the dots based on our structural characterizations.

The electronic and optical properties of carrier complexes confined in the QDs were calculated within the eight-band $\kp$ and configuration-interaction methods, including the expected range of energies for the neutral exciton and binding energies for the biexciton and charged exciton.
The optical properties of individual QDs were studied by high-spatially-resolution photoluminescence in mesa processed structures.
The experimentally obtained binding energies for the exciton and the biexciton were in a good agreement with the theoretically calculated values.
The close to symmetric shape of the QDs should result in near zero fine structure splitting.
However the measurements revealed FSS ${\sim(50\pm5)}~\si{\micro\electronvolt}$.
This, together with the increased carrier lifetime in mesa-processed structures compared to the planar structure, suggests that the defects introduced during the dry etching process can distort the QD electronic and optical properties from the average picture derived from the ensemble studies. 
Thus, further investigation and optimization of the processing are required.

Finally, the QDs show excellent single-photon emission properties: the Hanbury-Brown and Twiss-type interferometric experiment on a charged exciton revealed high purity of single-photon emission at $\SI{\sim1500}{\nano\meter}$ of $\mathcal{P}=(92.5 \pm 7.5)\%$, showing that these dots are promising candidates for single-photon emitters at the third telecom window.

\section{Methods}
\paragraph*{Fabrication.} The QDs were grown in the low-pressure MOVPE TurboDisc reactor using arsine (AsH$_3$), phosphine (PH$_3$), tertiarybutylphosphine (TBP) and trimethylindium (TMIn) precursors with H$_2$ as a carrier gas.
The growth sequence began with a $\SI{0.5}{\micro\meter}$-thick InP buffer layer deposited on an $(001)$-oriented InP substrate at $\SI{610}{\degreeCelsius}$.
Then, the temperature was decreased to $\SI{360}{\degreeCelsius}$ and stabilized under TBP.
The deposition of indium droplets occurred under the TMIn flow rate of $\SI{13}{\micro\mol\per\minute}$ with nominally $1.8$ ML-thick indium layer.
The indium droplets were annealed for $\SI{60}{\second}$ and the temperature was being raised to $\SI{550}{\degreeCelsius}$ during $\SI{130}{\second}$ under AsH$_3$ with the flow rate of $\SI{52.2}{\micro\mol\per\minute}$.
Afterwards, the temperature and AsH$_3$ flow were kept constant for $\SI{180}{\second}$ and finally the $\SI{30}{\nano\meter}$-thick InP was deposited. 
We repeated the indium droplet deposition and the annealing sequence for the surface QDs.
For the reference structure, without annealing, after the $\SI{60}{\second}$-long waiting time, the temperature was raised to $\SI{475}{\degreeCelsius}$ during $\SI{70}{\second}$ under AsH$_3$ ambient followed by the immediate InP layer deposition and the same sequence for the surface QD array finished the structure.
For the $\upmu$PL studies, the annealed structure was additionally processed to fabricate large mesa structures (size of $\SIrange{2}{3}{\micro\meter}$) in electron cyclotron resonance-reactive ion etching (RIE) in an Ar$^{+}$/Cl$^{-}$ plasma.

\paragraph*{Scanning transmission electron microscopy.}
Electron transparent lamellas of the cross-sections of the samples were prepared by focused ion beam milling (FIB) using a Helios Nanolab dual beam instrument.
The FIB milling and polishing was carried out using a Ga$^{+}$ beam at $\SI{30}{\kilo\volt}$ and with currents ranging from $\SI{2}{\nano\ampere}-\SI{20}{\pico\ampere}$.
The samples were then further polished in an Ar$^{+}$ Nanomill instrument at $\SI{700}{\volt}$ in order to remove FIB induced damage.
High-angle annular dark-field STEM images of the samples were acquired using a FEI Titan 80-300 instrument fitted with a field emission gun and with an aberration-corrector on the probe forming lenses, operated at $\SI{300}{\kilo\volt}$.
The electron probe had a convergence semi-angle of $\SI{18}{\milli\radian}$ and the images were recorded on an annular detector with an inner collection semi-angle of $\SI{50}{\milli\radian}$.
Geometric phase analysis was applied to the images to obtain lattice displacement maps along the $[001]$ direction using the freely available FRWRtools script~\cite{FRWRtools}.
The obtained changes in the lattice spacing along the growth direction and the Poisson ratio of InAs were then used to estimate the As content of the examined regions. 

\paragraph*{Optical experiments.} During the optical experiments, the structures were held in a helium-flow cryostat allowing for the sample's temperature control in the range of $\SIrange{4.2}{300}{\kelvin}$.
For the $\upmu$PL studies, the structures were excited by a $\SI{640}{\nano\meter}$ line from a cw semiconductor laser diode through a high-numerical-aperture ($\mathrm{NA}=0.4$) microscope objective with $20\times$ magnification.
The same objective was used to collect the $\upmu$PL signal and to direct it for the spectral analysis with a $\SI{1}{\meter}$-focal-length monochromator equipped with a liquid-nitrogen-cooled InGaAs multichannel array detector, providing spatial and spectral resolution of $\SI{\sim2}{\micro\meter}$ and   $\SI{\sim25}{\micro\electronvolt}$, respectively.
Polarization properties of emitted light were analyzed by rotating the half-wave retarder mounted before a fixed high-contrast-ratio ($10^6:1$) linear polarizer placed in front of the monochromator's entrance.
Autocorrelation histograms and TRPL were measured in a similar setup.
However, the structure was excited by a $\SI{787}{\nano\meter}$ cw laser line for the correlation spectroscopy, and by a train of $\SI{\sim50}{\pico\second}$-long pulses at the frequency of $\SI{80}{\mega\hertz}$, and the central photon wavelength of $\SI{805}{\nano\meter}$.
In this case, the collected photons were dispersed by a $\SI{0.3}{\meter}$-focal-length monochromator equipped either with the InGaAs multichannel array detector or NbN-based superconducting nanowire single-photon detectors with $\SI{\sim90}{\percent}$ efficiency in the 
$\SI{1.5}{\micro\meter}-\SI{1.6}{\micro\meter}$ range and $\sim200$ dark counts per second.
A multichannel picosecond event timer analyzes the single photon counts as the time-to-amplitude converter with the $\SI{256}{\pico\second}$ channel time bin width.
The overall temporal resolution of the setup is $\SI{\sim80}{\pico\second}$.

The degree of linear polarization (DOLP) is defined as 
\begin{equation}
    \mathrm{DOLP} = \frac{I_{\mathrm{max}}-I_{\mathrm{min}}}{I_{\mathrm{max}}+I_{\mathrm{min}}},
\end{equation}
where $I_{\mathrm{max}}$ ($I_{\mathrm{min}}$) is the maximal (minimal) PL intensity.
The quenching of the integrated $\upmu$PL intensity was fitted with the standard formula with a single activation processes:
\begin{equation}
I(T)=I_0/\left[ 1+B\exp{\left(-E_{\mathrm{A}}/k_{\mathrm{B}}T\right)} \right],
\end{equation}
where $I_0$ is the $\upmu$PL intensity for $T \to 0$, $E_{\mathrm{A}}$ is the activation energy, and $B$ is the quenching rate.
The autocorrelation histogram is fitted with the function of the form
\begin{equation}
C(\tau)=A\left[1-\left(1-g^{(2)}_{\mathrm{fit}}(0)\right) \exp{\left(-|\tau|/t_r\right)}\right],
\end{equation}
where $g^{(2)}_{\mathrm{fit}}(0)$ is the single-photon emission purity, $A$ in an average coincidence level, and $t_r = 1/(\Gamma+W_{\mathrm{P}})$ is the antibunching time constant with $\Gamma$ being the electron-hole radiative recombination rate, and $W_{\mathrm{P}}$ -- the effective pump rate.

\paragraph*{Calculation of electronic and optical properties.} The material composition profiles of simulated QDs, as presented in \subfigref{QDscheme}{a}, were discretized and represented on a numerical grid.
The strain field was calculated within the standard continuum elasticity theory, such that it minimizes the total elastic energy of the system.
As the materials are noncentrosymmetric, the shear components of the strain tensor lead to a nonuniform piezoelectric field~\cite{Caro2015}, which was calculated taking into account up to second-order terms in strain.
Next, electron and hole (time-reversed valence-band electron) eigenstates were calculated using a numerical implementation~\cite{Gawarecki2014} of the eight-band envelope-function $\kp$ method~\cite{Burt1992,Foreman1993} including the calculated strain and piezoelectric fields as well as spin-orbit effects (for the explicit form of the Hamiltonian and details of numerical implementation, see Ref.~\cite{Gawarecki2018}; for material parameters used for the InAs/InP material system see Ref.~\cite{Holewa2020PRB} and references therein).
Calculated eigenstates have the form of discretized envelope functions for each of the eight bands included in the model, which allows us to determine, e.\,g., the light-hole admixture to the hole ground state.
For each of simulated QDs a $40\times40$ electron-hole single-particle basis was computed and used to calculate the eigenstates of excitons, charged excitons and biexcitons within the configuration-interaction method by exact diagonalization of the electron-hole Coulomb and phenomenological anisotropic exchange interactions expanded in the configuration basis.
Using the light-matter coupling Hamiltonian in the dipole approximation~\cite{Thraenhardt2002}, optical transition dipole moments and resultant radiative lifetime and degree of polarization were calculated for each of the carrier-complex states.

\begin{acknowledgements}
We acknowledge support from the Danish National Research Foundation via Research Centre of Excellence NanoPhoton (ref. DNRF147).
P. H. was funded by the Polish National Science Center within the Etiuda 8 scholarship (Grant No. 2020/36/T/ST5/00511) and by the European Union under the European Social Fund.
V. G. D. gratefully acknowledges financial support of St. Petersburg State University under the research grant 75746688.
We are grateful to Krzysztof Gawarecki for sharing his implementation of the $\kp$ method. Numerical calculations have been carried out using resources provided by Wroclaw Centre for Networking and Supercomputing (https://wcss.pl). 
\end{acknowledgements}

\FloatBarrier

%

\input{SI}
\end{document}

%% file: SI.tex
\clearpage
\onecolumngrid

\begin{center}
\textbf{\large Supplemental Material: Droplet epitaxy symmetric InAs/InP quantum dots for quantum emission in the third telecom window: morphology, optical and electronic properties}
\end{center}
\setcounter{equation}{0}
\setcounter{figure}{0}
\setcounter{table}{0}
\setcounter{page}{1}
\setcounter{section}{0}
\renewcommand{\thesection}{S-\Roman{section}}
\makeatletter
\renewcommand{\theequation}{S\arabic{equation}}
\renewcommand{\thefigure}{S\arabic{figure}}
\renewcommand{\thetable}{S\arabic{table}}

\section{EDX data}\label{Sec:EDX-data}

In this section we show the raw EDX data which form the basis for the EDX profiles plotted in \subfigref{TEM}{g} for the QD and \subfigref{TEM}{i} for the 2D layer.

\begin{figure*}[h!]
 \centering
 \includegraphics[width=0.75\linewidth]{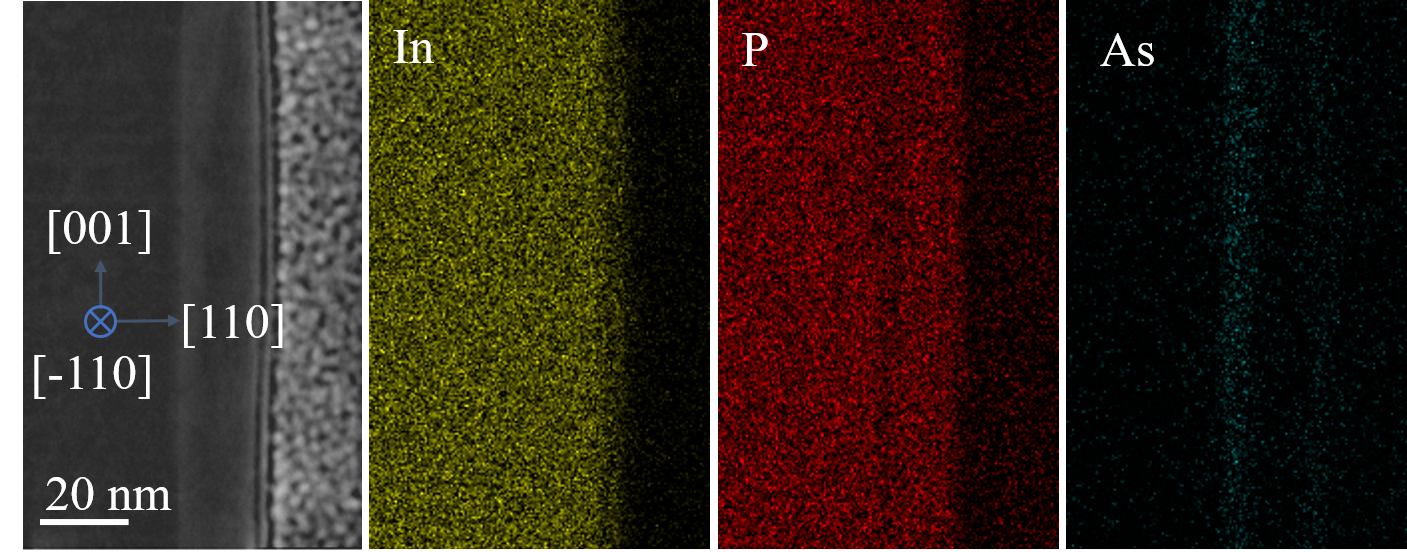}
 \caption{EDX data for a QD viewed along the [-110] direction.}
 \label{fig:EDX-QD}
\end{figure*}

\begin{figure*}[h!]
 \centering
 \includegraphics[width=0.75\linewidth]{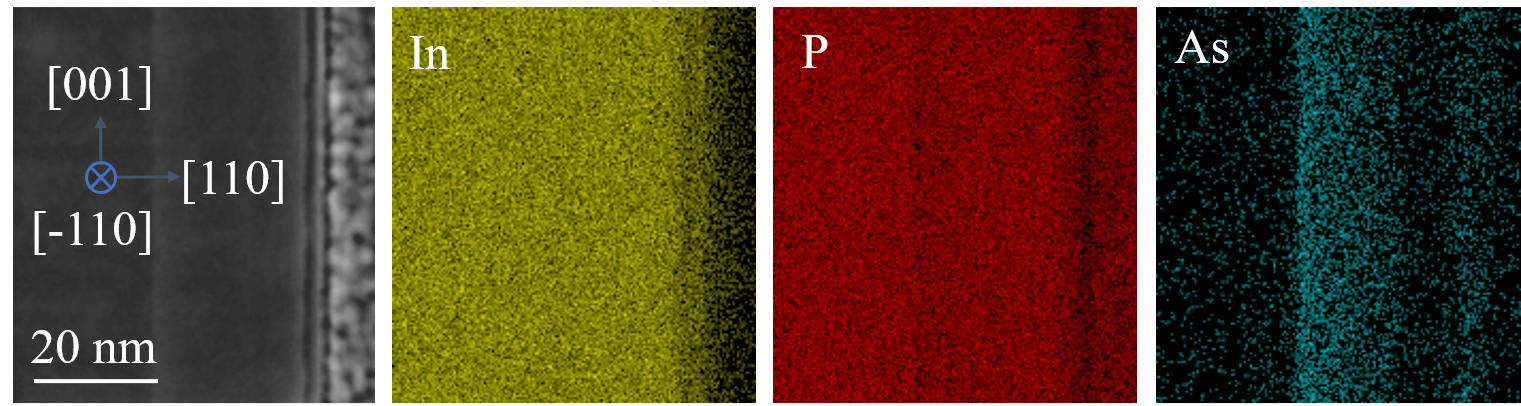}
 \caption{EDX data for the 2D layer viewed along the [-110] direction.}
 \label{fig:EDX-2D-layer}
\end{figure*}

\section{Results of the pit etching models.}
\begin{figure*}[h!]
 \centering
 \includegraphics[width=0.75\linewidth]{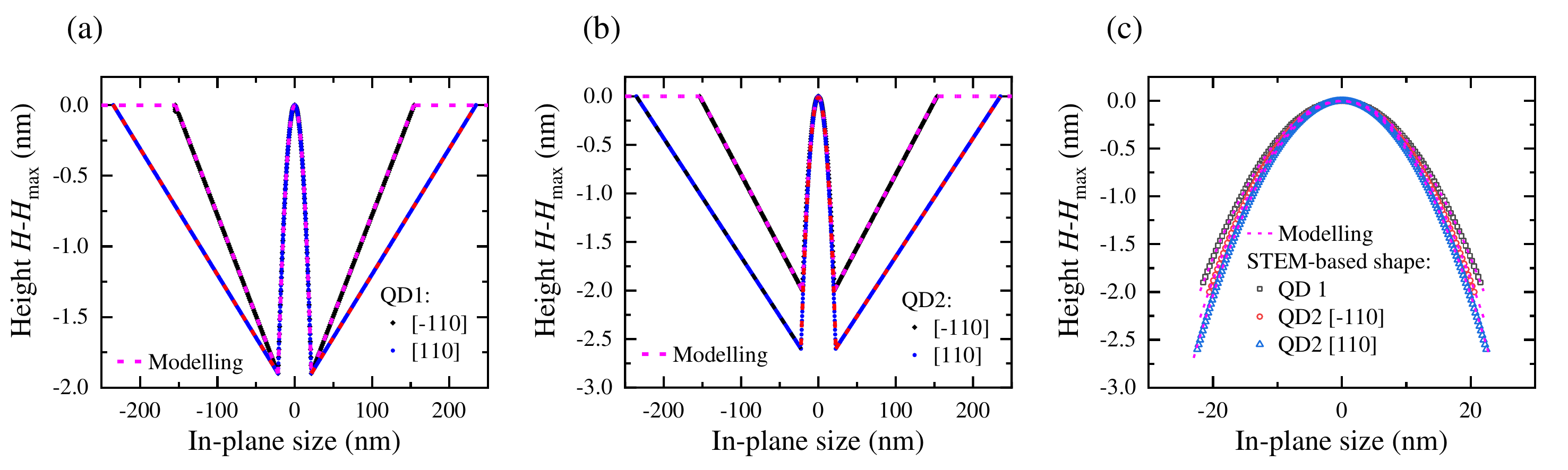}
 \caption{Comparison of the assumed geometry of the pits and the 2D layer with the results of kinetic theory modelling for (a)~the symmetric QD (QD1),
 (b)~realistic QD with a slight asymmetry (QD2).
 (c)~The close-up for the bottom of the QD.}
 \label{fig:Nucleation-theory}
\end{figure*}

The results of QD-induced etching of InP are summarized in \figref{Nucleation-theory}.
Assumed geometry of the 2D layer is shown with the solid lines while the dashed lines show the results of the model, described in the text.

\newpage
\section{Calculation of excitonic complexes in QDs}

\begin{figure}[!h]
 \centering
 \includegraphics[width=0.3\linewidth]{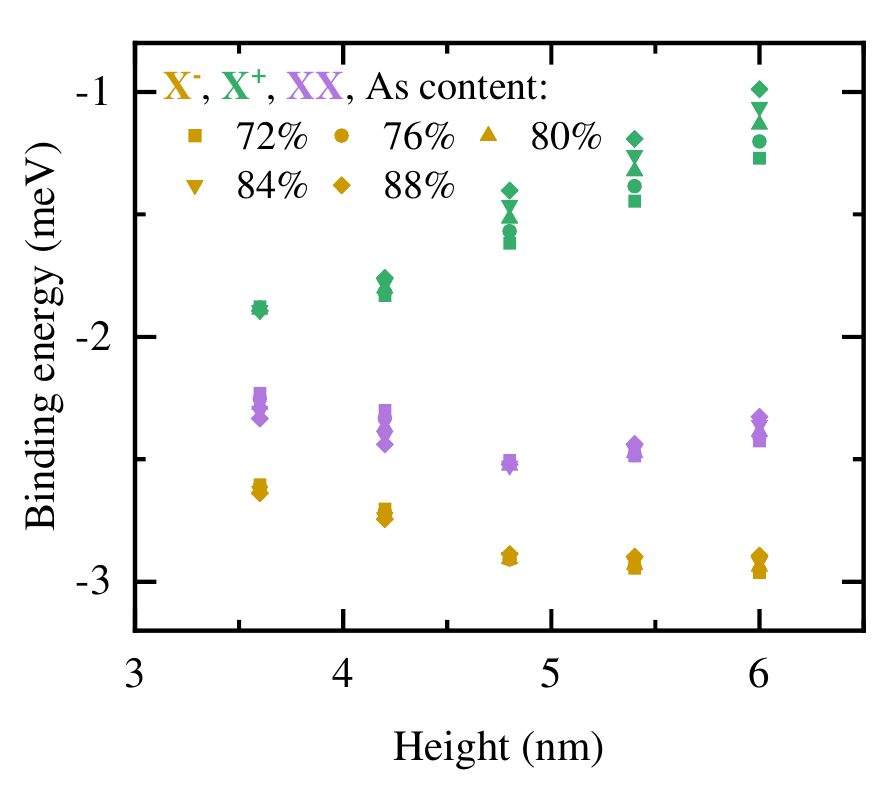}
 \caption{Calculated binding energies for excitonic complexes confined within QDs as a function of QD height and As content.}
 \label{fig:calc-binding-en}
\end{figure}

\figref{calc-binding-en} presents the results of binding energy calculations for a broad range of QD composition (As content, $x=\SIrange{72}{88}{\percent}$) and QD height ($H=\SIrange{3.6}{6}{\nano\meter}$).

